\documentclass{jpp}

\usepackage{graphicx}
\usepackage{amsmath,amssymb}
\usepackage{xcolor}

\usepackage[utf8]{inputenc}
\usepackage[T1]{fontenc}
\usepackage{amsmath}
\usepackage{hyperref}
\usepackage{soul}
\usepackage{booktabs}
\usepackage{array}

\DeclareMathOperator{\sym}{sym}

\newcommand{\tensor}[1]{\mathrm{#1}}

\renewcommand{\vec}[1]{\mathbf{#1}}

\newcommand{\unitvec}[1]{\hat{\vec{#1}}}
\newcommand{\muphyii}{\textit{muphyII}~}

\shorttitle{Optimal Landau-type closure parameters for two-fluid simulations}
\shortauthor{S. Lautenbach, J. Lübke, M.E. Innocenti, K. Kormann and R. Grauer}

\title{Optimal Landau-type closure parameters for two-fluid simulations of plasma turbulence at kinetic scales}

\author{Simon Lautenbach\aff{1}\corresp{\email{simon.lautenbach@ruhr-uni-bochum.de}},
  Jeremiah Lübke\aff{1},
  Maria Elena Innocenti\aff{1},
  Katharina Kormann\aff{2},
  \and Rainer Grauer\aff{1}}

\affiliation{
\aff{1}Theoretical Physics I, Ruhr University Bochum, Universitätsstraße 150, D-44801 Bochum, Germany
\aff{2}Numerical Mathematics, Ruhr University Bochum, Universitätsstraße 150, D-44801 Bochum, Germany
}

\begin{document}

\maketitle

\begin{abstract}
Two fluid simulations using local Landau-fluid closures derived from linear theory provide an efficient computational framework for plasma modelling, since they bridge the gap between computationally intensive kinetic simulations and fluid descriptions. Their accuracy in representing kinetic effects depends critically on the validity of the linear approximation used in the derivation: the plasma should not be too far from local thermodynamic equilibrium, LTE. However, many of the problems where these models are of particular interest (such as plasma turbulence and instabilities) are in fact quite far from LTE. The question then arises, if kinetic scale processes are still sufficiently well captured outside of the theoretical regime of applicability of the closure. In this paper, we show that two fluid simulations with Landau fluid closures can effectively reproduce the energy spectra obtained with fully kinetic Vlasov simulations, used as references, as long as the local closure parameter is appropriately chosen. Our findings validate the usage of two fluid simulations with Landau-fluid closure as a possible alternative to fully kinetic simulations of turbulence, in cases where being able to simulate extremely large domains is of particular interest.
\end{abstract}

\section{Introduction}

Space and astrophysical plasmas are characterized by the extreme separation between system scales, where large-scale drivers act, and kinetic scales, where wave/particle interaction processes result into particle heating acceleration and energy dissipation. 
The interaction between system and kinetic scales is not trivial, as we highlight in the three examples that follow: First, in the terrestrial magnetosphere, magnetotail reconnection~\citep{zweibel2009magnetic, yamada2010magnetic} is driven by the global-scale Dungey cycle~\citep{dungey1965length},  triggered by magnetic field lines opening and reconnecting at the electron scales~\citep{kuznetsova2001collisionless},  
and results into global reshaping of the magnetospheric field, as well as particle heating and acceleration~\citep{angelopoulos2013electromagnetic}.
Second, energy injection and dissipation scales in turbulence are separated by several decades in wavenumber and frequency~\citep{frisch1995turbulence, bruno2013solar}. Intermediate scales are however of extreme relevance, since meso-scale processes lead to the formation of structures, such as elongated reconnecting current sheets, which fundamentally influence how energy flows across the scales~\citep{sundkvist2007dissipation}. 
Third, in the solar wind, plasma expansion in interplanetary space and the resulting variation in background plasma parameters trigger ion and electron instabilities, which in turn contribute to constrain large scale solar wind properties such as temperature anisotropies, plasma betas, heat flux~\citep{matteini2013signatures, vstverak2008electron, micera2021role, verscharen2022electron, micera2025quasi}.

For this reason, one of the main challenges in plasma modelling is to capture, in a single simulation, the non-trivial interaction between small (kinetic) and large (system) scales. 
Fully kinetic models, i.e., Vlasov~\citep{cheng1976integration} or Particle In Cell (PIC)~\citep{hockney1988computer} methods, model kinetic scale processes self-consistently and without approximations on the state of the plasmas by discretizing the Vlasov equations for particle evolution.
They do so either directly in velocity space or by sampling the distribution function using computational particles each representing a number of point particles inhabiting nearby regions of phase space, respectively.
However, their computational costs are often prohibitive for the study of physical systems of interest, whose sizes very often exceed characteristics kinetic scales of several order of magnitude.
Methods can be devised to reduce the computational costs of both Vlasov and PIC simulations, including semi-implicit~\citep{hewett1987electromagnetic, vu1992celest1d, lapenta2017multiple,kormann2021} and fully implicit~\citep{chen2011energy, markidis2011energy} dicretizations, adaptive grids~\citep{fujimoto2008electromagnetic, vay2004application, innocenti2013multi}, low-rank tensor compression \citep{kormann2015, einkemmer2018, allmann-rahn-grauer-kormann:2021,ye-loureiro:2024,einkemmer2024}, asymptotic-preserving schemes \citep{degond2010asymptotic,ji2023asymptotic}, and ad-hoc techniques addressing specific characteristic of the system such as expansion or contraction~\citep{riquelme2012local, ahmadi2017simulation, innocenti2019semi}.
This notwithstanding, the computational cost of large-scale kinetic simulations requires that reduced models have to be employed.

At the opposite extreme of the computational cost versus physical accuracy, one finds MagnetoHydroDynamics (MHD).
The MHD description makes a number of implicit assumptions on the state of the plasma, including quasi-neutrality, massless electrons (the plasma is considered as a single fluid), only ideal electric field components, and scalar pressure~\citep{ledvina2008modeling}.
As long as one is not interested in processes that require violating one of these implicit assumption, MHD models can be used for relatively cheap large scale simulations including, e.g., global heliospheric simulations~\citep{gombosi2018extended, poedts2020european}, which are too expensive for practically any other model.
If one aims for simulations that can tackle larger boxes with respect to fully kinetic simulations, but include more physical details than MHD simulations, two types of possibilities present themselves. 
On the one hand, MHD and fully kinetic (most often, in the PIC flavour) codes can be coupled, under the assumption that the MHD  assumptions are broken only in a small fraction of the domain, where the kinetic description is needed.
Coupled MHD/PIC codes are used in a number of environments (such as the terrestrial~\citep{lapenta2016multiscale, chen2017global}  and planetary~\citep{chen2019studying} magnetospheres and the solar corona~\citep{haahr2025coupling}) where system scale drivers result in kinetic scale processes, often magnetic reconnection. 
On the other hand, intermediate models between fully kinetic and MHD can be used, which require a reduced number of implicit assumptions on the state of the plasma with respect to MHD.
Table 1 in~\citet{ledvina2008modeling} details the implicit assumptions made in multi-fluid, Hall-MHD and in hybrid models. 
Multi-fluid multi-moment models describe plasmas as composed of multiple fluids, and evolve a varying number of their velocity moments (density, current and energy in the 5-moment model; density, current and the pressure tensor in the 10-moment model), as we describe in detail in section~\ref{sec:muphy}.

The \muphyii{}framework~\citep{lautenbach-grauer:2018, allmannrahn-lautenbach-deisenhofer-grauer:2024} used in this work implements both a Vlasov and multi-fluid (5 and 10 moment) descriptions, which can be used either stand-alone or in coupled mode. In this regard, \muphyii{}is a peculiarity among plasma modelling frameworks, since one can use both reduced descriptions and code coupling to extend domain size and physical duration of plasma simulations.

In this paper we benchmark 10-moment simulations of plasma turbulence against their fully kinetic, Vlasov counterparts.
We focus on the 10-moment model because plasma turbulence is a typical multi-scale plasma process, where one wants to simulate both very large domains (hence, fluid models are preferable to fully kinetic models, due to their lower computational cost), as well as the scales and mechanism of energy dissipation via wave-particle interactions.
This transfer process, fundamentally a kinetic phenomenon, plays a crucial role in determining the shape of energy spectra, particularly at scales intermediate between the ion and electron scales. For this reason, using a high number of velocity moments and a good choice of \textit{closure} of the moment hierarchy is of particular importance.
Landau fluid closures, first introduced by \citet{hammett-perkins:1990}, approximate linear kinetic effects within a fluid representation, particularly for parallel heat transport, by introducing specific relations between higher- and lower-order moments of the distribution function \citep[see also][]{hunana2019introductory}. 
They are derived in regimes where linear approximation holds, meaning the velocity distribution function is only slightly perturbed with respect to the Maxwellian distribution. However, in this work, we show that this approximation can still be used far from equilibrium, by adjusting closure parameters.

In the local (as opposed to non-local,~\citet{ng2017simulations}) three-dimensional generalization of the Hammett-Perkins closure in weakly magnetized plasmas, a free parameter $k_0$ emerges that represents a characteristic wave number related to the maximum damping rate~\citep{wang2015comparison}. \citet{allmann-rahn-trost-grauer:2018} and \citet{ng2020improved} have shown that 10-moment simulations closed with different flavors of Landau fluid closures reproduce well fully kinetic magnetic reconnection results.
10-moment models with Landau fluid closures have also delivered favourable results in reproducing pressure-gradient and pressure-anisotropy driven instabilities, such as the Lower Hybrid Drift Instability~\citep{ng2020improved, allmann-rahn-lautenbach-grauer-etal:2021} and the firehose instability~\citep{walters2024electron}.
It is often the case that the free parameter $k_0$ that delivers the best comparison with fully kinetic simulations is related to characteristic scales of the plasma species~\citep{ng2020improved}.

The goal of our 10-moment turbulence simulations is not to perfectly reproduce the kinetic damping and energy transfer process, which would require non-local closures and possibly a higher number of moments, but rather to identify if a closure parameter exists, that can reproduce sufficiently well the \textit{global} effects of energy dissipation at kinetic scales, i.e. the slope and shape of the energy spectra.
We will do so via a rather brute force approach: after identifying a suitable Landau fluid closure, i.e., the one described and validated in~\citet{allmann-rahn-lautenbach-grauer-etal:2021}, we will run multiple 10-moment turbulence simulations with varying values of the $k_0$ parameters. We will then identify as ``optimal'' $k_0$ parameters, for both ions and electrons, the ones which deliver turbulent spectra which better compare with the Vlasov results.  

We proceed in two distinct phases: First, we establish an optimal electron closure parameter by comparing hybrid simulations (kinetic ions, fluid electrons) against fully kinetic reference simulations, focusing on matching energy spectra across the broadest possible range of scales.
Second, we apply this optimized electron closure in full fluid simulations while varying the ion closure parameter to determine its optimal value. We compare and comment simulation results across different models (Vlasov, hybrid, two fluid) and with different parameter choices.

This paper is organized as follows. First, in section \ref{sec:muphy}, we describe the so-called closure problem for 10-moment models. In section \ref{sec:setup}, we describe the three types of problems that we tackle in this paper: Landau damping, Kelvin-Helmholtz Instability and finally, decaying turbulence simulations. While our final aim is decaying turbulence simulations, we first address ``simpler'' problems such as Landau damping and Kelvin-Helmholtz simulations because they allow us to tackle, in isolation, processes that will occur, together, in turbulence simulations. Using the Landau damping test case, we show that an optimal value of the $k_0$ parameter can be identified, allowing a fluid description to capture this intrinsically kinetic process. Naturally, this comes in reduced form, leading to discrepancies compared to fully kinetic simulations.
With the Kelvin-Helmholtz simulations we demonstrate that 10-moment models can satisfactorily reproduce kinetic results for velocity shear instabilities, which are expected to develop in the decaying turbulence simulation. Parameter optimization and model comparison for our three test cases of interest are presented in section~\ref{sec:res}. Discussion and Conclusions follow in sections~\ref{sec:discussion}.

\section{Theoretical Framework} \label{sec:muphy}

The behaviour of collisionless plasmas is governed by the evolution of the distribution function $f_s(\vec x, \vec v, t)$ for each species $s$, as described by the Vlasov equation:
\begin{equation}\label{eq:vlasov}
\partial_t f_s +\vec{v} \cdot \nabla_x f_s + \frac{q_s}{m_s} (\vec E + \vec v \times \vec B) \cdot \nabla_v f_s = 0
\end{equation}
where $\vec E$ and $\vec B$ are the electric and magnetic field respectively, $q_s$ and $m_s$ are the charge and mass of species $s$.  This equation couples to Maxwell's equations, which determine the self-consistent electromagnetic fields:
\begin{equation}\label{eq:maxwell}
\partial_t \vec B = -\nabla \times \vec E,\quad
\partial_t \vec E = c^2 \left(\nabla \times \vec B - \mu_0 \vec J\right),\quad
\nabla \cdot \vec B = 0,\quad
\nabla \cdot \vec E = \frac{\rho_c}{\varepsilon_0},
\end{equation}
where $\rho_c$ represents the charge density,  $\vec J$ the current density, and $c$, $\mu_0$, and $\varepsilon_0$ are the speed of light, vacuum permeability, and permittivity, respectively.

The so-called 10-moment model \citep{hakim:2008,johnson-rossmanith:2010} is a suitable compromise between reduced computational cost and expressivity of the description. The 10-moment model includes the first ten fluid moments which are obtained by integrating over velocity space: the particle density $n_s = \int f_x(\vec x, \vec v, t) \, \text{d} \vec v$, mean velocity $\vec u_s = \int \vec v f(\vec x,\vec v, t) \, \text{d} \vec v/n_s$ and the momentum flux density $\mathcal{P}_{s} = m_s \int \vec v \otimes \vec v f_s(\vec x, \vec v, t) \, \text{d} \vec v$. The equations are then obtained from integration of the Vlasov equation as follows:
\begin{align}
\frac{\partial n_{s}}{\partial t} &+ \nabla \cdot (n_{s} \mathbf{u}_{s}) = 0,\\
m_{s} \frac{\partial (n_{s} \mathbf{u}_{s}) }{\partial t} 
&- n_{s} q_{s} (\mathbf{E} + \mathbf{u}_{s} \times \mathbf{B}) + \nabla \cdot \mathcal{P}_{s} = 0,\\
\frac{\partial \mathcal{P}_{s}}{\partial t}
&- q_{s} (n_{s} \text{sym}[\mathbf{u}_s \mathbf{E}]
+ \frac{1}{m_{s}} \text{sym}[\mathcal{P}_{s} \times \mathbf{B}])
  + \nabla \cdot \mathcal{Q}_{s} = 0. 
\label{eq:fluid}
\end{align}
Here we denote the generalized tensor product by $(\tensor a\times\vec b)_{ijk} = \epsilon^{klm}a_{ijl}b_m$, the standard outer product by $\otimes$, and $\sym[\cdot]$ represents the symmetrization operator $\sym[\tensor a]_{ijk} = a_{ijk}+a_{jki}+a_{kij}$. The last equations involves the next higher moment, the third moment $\mathcal{Q}_{s} = \int \vec v \otimes \vec v \otimes \vec v f(\vec x, \vec v, t) \, \text{d} \vec v$, which is related to the heat flux  $\mathrm{Q}_{i j k}$ as
\begin{equation}
\mathrm{Q}_{s}=\mathcal{Q}_{s}-\operatorname{sym}[\mathbf{u} \otimes \mathcal{P}]_{s}+2 m_s n_s \mathbf{u}_{s} \otimes \mathbf{u}_{s} \otimes \mathbf{u}_{s} .
\label{eq:thirdMom}
\end{equation}

The critical step in developing a practical fluid model lies in the closure of the moment hierarchy.

The simplified 5-moment model is obtained when assuming isotropic pressure and zero heat flux. In this case, the tensorial pressure reduces to the energy density $\mathcal{E}_s = \frac{m_s}{2} \int |\vec{v}|^2 f_s(\vec x , \vec v, t) \, \text{d}\vec v$ and the fluid equations are given by
\begin{align}
\partial_t n_s &+ \nabla \cdot (n_{s} \vec u_s) = 0\\
m_{s} \partial_t (n_s \vec u_s) &- n_s q_s (\vec E + \vec u_s \times \vec B)
+ \frac{1}{3}\nabla \left( 2 \mathcal{E}_s - m_s n_s |\vec{u}_s|^2 \right) + \nabla \cdot \left( m_s n_s \vec{u}_s \otimes \vec{u}_s \right)= 0\\
\partial_t  \mathcal{E}_s &- q_s n_s \vec{u}_s \cdot \vec{E} + \frac{1}{3} \nabla \cdot \left( \vec{u}_s (5\mathcal{E}_s - m_sn_s |\vec{u}_s|^2\right) = 0.
\label{eq:fluid5}
\end{align}

This 5-moment model represents the simplest fluid description and completely neglects kinetic effects.
In the case of a 10-moment model, one needs to find a suitable formulation for the divergence of the heat flux tensor in Eq.~\ref{eq:thirdMom}.

We use here a local~\citep{sharma2006shearing} closure that receives information from the temperature gradient,
\begin{equation}\label{eq:tenmoment_closure}
\nabla \cdot Q_s = -\frac{\chi}{k_{s,0}}\,n_s\,v_{\text{th},s}\,\nabla^2\,\tensor T_s\,.
\end{equation}

With respect to other closures based directly on pressure tensors, the temperature-gradient approach offers better numerical stability during adaptive coupling between kinetic and fluid regions~\citep{allmann-rahn-trost-grauer:2018,allmann-rahn-lautenbach-grauer:2022}, making it particularly suitable for multi-scale simulations with localized kinetic effects. It introduces the parameter $1/k_{s,0}$ which has units of length and can be interpreted as a characteristic scale for heat transport processes. The dimensionless factor $\chi$ is typically set to $\sqrt{8/\pi}$ based on the linear theory of Landau damping \citep{hammett-perkins:1990}, while $k_{s,0}$ remains a free parameter that must be calibrated to match the desired kinetic response. Throughout this paper, unless otherwise specified, we normalize $k_0$ to the skin depth of the respective particle species, i.e., electron, $d_e$, or ion, $d_i$, skin depth. The optimal value depends on the dominant physical processes in the plasma regime under consideration.

It is interesting to examine the relation of the 10-moment model with the 5-moment and (by extension) MHD limits. Unlike some fluid models that can formally reduce to either under appropriate limiting conditions, our Landau-fluid closure for the 10-moment model cannot inherently recover the 5-moment limit for any choice of the free parameter $k_{s,0}$. This occurs because the parameter creates a mathematical tension between two incompatible conditions: as $k_{s,0}$ decreases (or $1/k_{s,0}$ increases), the heat flux term in equation \eqref{eq:tenmoment_closure} becomes stronger, effectively enhancing thermal diffusion and driving the distribution function more aggressively toward an isotropic and thus Maxwellian state. However, the five-moment model assumes zero heat flux rather than enhanced heat transport. Conversely, as $k_{s,0}$ increases (or $1/k_{s,0}$ decreases), the heat flux contribution diminishes, potentially allowing non-isotropic distributions to persist, another deviation from the 5-moment model, which fundamentally assumes local thermodynamic equilibrium. This inherent tension underscores that our closure represents an intermediate regime between full kinetic physics and 5-moment models, capturing important kinetic effects while retaining computational efficiency, but without the ability to formally reduce to either limiting case through parameter adjustment alone.

\section{Simulation Setup}\label{sec:setup}

We study three test cases, Landau damping, the Kelvin--Helmholtz instability, and a decaying turbulence test case.  We will study the influence of the choice of $k_0$ in the 10-moment approximation.  
The distinct models  employed in this study are:
\begin{enumerate}
\item[(a)] Fully kinetic Vlasov simulation of both species: high fidelity reference;
\item[(b)] Hybrid model with fully kinetic ions, 10-moment electrons: next model in the hierarchy to separate the influence of the moment model approximation on the two species;
\item[(c)] 10-moment model for both species;
\item[(d)] 10-moment ions, 5-moment electrons;
\item[(e)] 5-moment models for both species (with isotropic adiabatic closure).
\end{enumerate}

\begin{table}
\centering
\begin{tabular}{l@{\hskip 10pt}ccccc}
Simulation & Fully Kinetic & Hybrid & 10-moment & 10m/5m & 5-moment \\[5pt]
Landau Damping & \checkmark & \checkmark &  &  &  \\
Kelvin-Helmholtz & \checkmark & \checkmark & \checkmark & \checkmark & \checkmark \\
  decaying turbulence
           & \checkmark & \checkmark & \checkmark &  &  \\
\end{tabular}
\caption{Models employed across different simulation cases. Each cell indicates whether a particular model (column) was used for a specific simulation setup (row) for both the electron and ion components.}
\label{tab:model-usage}
\end{table}

Table~\ref{tab:model-usage} describes which model is used for the three simulation test cases described in the following sections.

The numerical implementation uses the \muphyii framework \citep{allmannrahn-lautenbach-deisenhofer-grauer:2024} with efficient algorithms chosen for their conservation properties and numerical stability. For the Vlasov equation, we utilize the positive flux-conservative (PFC) scheme developed by \cite{filbet-sonnendruecker-bertrand:2001}, enhanced with the backsubstitution method \citep{schmitz-grauer:2006} and a moment-matching algorithm that significantly improves energy conservation \citep{allmann-rahn-lautenbach-grauer:2022}. The electromagnetic fields evolve according to Maxwell's equations through the finite difference time domain (FDTD) method on a staggered grid, following the classical approach of \cite{yee:1966}.

The fluid models employ a central weighted essentially non-oscillatory (CWENO) reconstruction scheme \citep{kurganov-levy:2000}, combined with third-order Runge-Kutta (RK3) time integration \citep{shu-osher:1988}. To manage the difference in temporal scales between electron and ion dynamics, we implement electron subcycling within each ion timestep, improving computational efficiency while maintaining numerical stability.

\subsection{Landau Damping}

To investigate electron Landau damping, we initialize a one-dimensional periodic system with a small-amplitude electron density perturbation. The simulation domain spans $L_x = 4\pi\lambda_{D,e}$ where $\lambda_{D,e}$ is the electron Debye length, chosen to accommodate a single wavelength of the perturbation. The initial condition consists of a Maxwellian electron velocity distribution with density modulation:
\begin{equation}
    f_e(x,v,t=0) = n_0\sqrt{\frac{m_e}{2\pi\,T_{0,e}}}\left(1 + \alpha\cos(k_px)\right)\exp\left(-m_e\frac{v^2}{2T_{0,e}}\right)
\end{equation}
where $\alpha = 0.01$ ensures operation in the linear regime, and $k_p = 2\pi/L_x$ is the wave number of the perturbation. The background ion population remains uniform and stationary, providing a neutralizing background.  Space is resolved with $256$ cells, the ion velocity grid spans over $\pm 6\sqrt{2} \,\lambda_{D,e}\omega_{p,e}$, with $\omega_{p,e}$ the electron plasma frequency,  and is resolved with $256$ cells. We use a realistic mass ratio $m_i/m_e=1836$ and the temperature ratio is set to $T_i/T_e=1$.

\subsection{Kelvin-Helmholtz Instability}

The Kelvin-Helmholtz instability is initiated through a velocity shear layer in both ions and electrons in a doubly periodic domain of size $L_x \times L_y = 18 L_0 \times 27 L_0= 900 \; d_e \, \times \, 135  \, d_e$, where $L_0$ is the half-width of the initial shear layer. The shear layer extents along the $y$ direction and is located at the center of the box along $x$. At the boundary layer, together with the velocity shear, we initialize gradient of density and out-of-plane magnetic field following hyperbolic tangent profiles, maintaining pressure balance across the shear layer. To trigger the instability we impose a small perturbation on the ion velocity with $N=10$ modes according to
\begin{equation}
\delta \vec u = \hat{\vec e}_z \times \nabla \Psi \quad\text{with}\quad 
\Psi = \frac{\epsilon}{N} \exp\left[-\left(\frac{x-x_0}{L_0}\right)^2\right] \sum_{m=1}^N \frac{1}{m} \cos\left[2\pi m \frac{y}{L_y} + \phi_m\right]
\end{equation}
where $\epsilon$ is the perturbation amplitude and $\phi_m$ the individual phase amplitude.  We use $2400 \times 1600$ cells in $x$ and $y$, respectively, velocity is resolved with $48^3$ cells and spans $\pm 1 \,v_A$ for ions and $ \pm 2.3 \,v_A$ for electrons, with $v_A$ the Alfv\'{e}n speed, the mass ratio between ions and electrons is $m_i/m_e = 25$, the ratio between the initial ion and electron temperature is $T_{0i}/T_{0e} = 5$ and $\omega_{p,e}/\omega_{c,e} = 4$, with $\omega_{c,e}$ being the electron cyclotron frequency.

\subsection{Decaying Turbulence}

To initiate decaying two-dimensional turbulence with low velocity-magnetic field correlation, we adopt a Biskamp-Welter vortex configuration \citep{biskamp-welter:1989}. The initial magnetic and velocity fields are given by:

\begin{align}
\vec B   &= \frac13 \delta u\ \unitvec{e}_z\times\left(\cos\left[(2x+2.3)\frac{2\pi}{L}\right]+\cos\left[(y+4.7)\frac{2\pi}{L}\right]\right) + B_0\unitvec{e}_z\\
\vec u_s  &=         \delta B\ \unitvec{e}_z\times\left(\cos\left[ (x+1.4)\frac{2\pi}{L}\right]+\cos\left[(y+0.5)\frac{2\pi}{L}\right]\right) + u_{z,s}
\end{align}
The out-of-plane electron velocity $u_{z,e}$ is determined by the MHD current condition $\vec J = \nabla\times{\vec B}/\mu_0$, while maintaining $u_{z,i}=0$ for ions. The initial electrostatic field follows from ideal Ohm's law, $\vec E_0 = \vec u_i \times B_0\unitvec{e}_z$, with corresponding electron density perturbation $n_e = n_0 - \vec E_0 \varepsilon_0$.

Our parameter choices follow the benchmark case \texttt{A1} from \cite{groselj:2017}. In particular, we use equal electron and ion temperatures $T_e/T_i=1$, a low plasma beta of $\beta_s=0.1$, reduced mass ratio $m_e/m_i = 1/100$, reduced vacuum speed of light $c=18.174 v_{\text A}$ and a moderate initial fluctuation amplitude of $\varepsilon = \delta B/B_0 = \delta u/v_{\text A} = 0.2$.

The periodic computational box for the Biskamp-Welter turbulence setup extends to $L \times L = 8\pi d_i \times 8\pi d_i$. This domain is resolved with $1024 \times 1024$ grid points in configuration space, providing sufficient resolution to capture structures down to sub-ion scales while maintaining reasonable computational costs. Velocity space spans $\pm 16\,v_A$ for electrons and $\pm 1.6\,v_A$ for ions in each direction, resolved with $20^3$ cells.

\section{Parameter Optimization and Model Comparison}
\label{sec:res}

In this section, we will compare several reduced 10-moment simulations with different values of the $k_0$ parameter in equation~(\ref{eq:tenmoment_closure}) with their fully kinetic Vlasov counterparts. In each case, we will identify the optimal $k_0$ for that specific problem, and strive to relate it to characteristic plasma parameters, for the sake of generalization to different simulation setups. Each subsection addresses a different aim: section \ref{sec:landau-damping} examines whether there exists at all an optimal value of $k_0$ that best characterises the damping behaviour of the heat flux. Then, we proceed to investigate if the closure is able to reproduce specific plasma dynamics, i.e. plasma mixing processes (section~\ref{sec:khi}). Finally, we tackle our problem of interest, decaying turbulence, section~\ref{sec:biskamp-welter}.

\subsection{Landau Damping}\label{sec:landau-damping}

We begin our investigation with the electron Landau damping, a fundamental kinetic process that represents the archetypal wave-particle interaction mechanism in collisionless plasmas~\citep{nicholson1983introduction}. This phenomenon presents an ideal starting point for our investigation for several reasons. First, as a linear process with well-established theoretical understanding, it provides a clean benchmark against which we can evaluate fluid closures designed specifically to capture such kinetic effects. Second, the one-dimensional nature of the problem isolates the physics of interest without the complications of multi-dimensional coupling. Third, and most importantly, Landau damping represents the primary mechanism through which electromagnetic fluctuations transfer energy to particles in turbulent collisionless plasmas -- precisely the physics our optimized closure needs to capture correctly. 

\begin{figure}
  \centering
  \includegraphics[width=.6\textwidth]{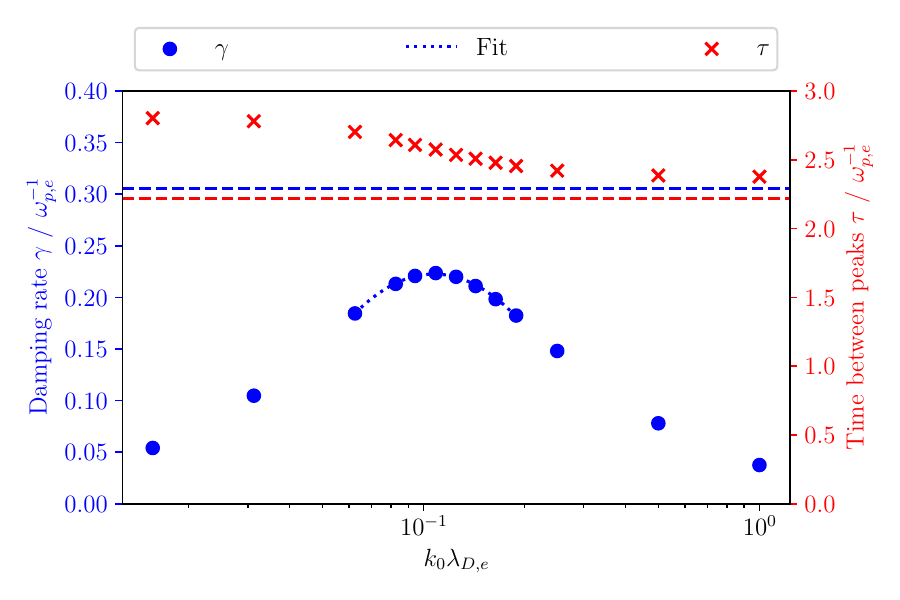}
  \caption{Landau damping in hybrid (Vlasov ion, fluid electron) simulations as a function of $k_0 \lambda_{D,e}$. The damping rate $\gamma$ is plotted in blue on the left axis, the half period of the electric field oscillations in red on the right axis. The dashed lines at the top and bottom are results from the fully kinetic benchmark simulation. A parabolic fit of the damping rate over the highest values (dotted blue) gives the closest approach to the benchmark value at $k_0=0.11\,\lambda_{D,e}^{-1}$ with a damping rate of $0.22\,\omega_{p,e}^{-1}$.}
  \label{fig:landau-damping}
\end{figure}

We conduct a systematic series of electron Landau damping simulations with a fluid treatment of electrons while varying the closure parameter $k_{0,e}$. The ions are simulated kinetically.

The resultant damping rate $\gamma$ and the half-period $\tau$ of the electric field oscillations (time between peaks in the absolute value of the electric field) are presented in figure \ref{fig:landau-damping}, on the left (blue) and right (red) axis respectively, as a function of the $k_0$ value. Results from the benchmark Vlasov simulations are depicted as a dashed line.

The observed behaviour of the damping rate $\gamma$ and half-period $\tau$ as functions of $k_0$ can be understood by considering the heat flux closure's physical interpretation. At very small $k_0$ values, the heat flux becomes very large, causing rapid thermalization that overwhelms the wave-particle resonance responsible for Landau damping. Conversely, at large $k_0$ values, the heat flux contribution becomes negligible, essentially decoupling the higher-order moments from the lower-order dynamics and preventing effective energy transfer between waves and particles. The optimal value represents a balance where the closure-induced dissipation most accurately mimics the kinetic resonance condition that drives Landau damping in the fully kinetic description.

Among the various $k_0$ values tested, we observe that $k_0 \lambda_{D,e}  \approx 0.11$ produces a damping rate that most closely approximates the kinetic value. The oscillation frequency shows less sensitivity to parameter variation but still exhibits reasonable correspondence near this optimal value. We select this optimal value based primarily on the damping rate, as it directly reflects the efficiency of energy transfer between fields and particles, a critical process for accurately modelling turbulent cascades in more complex scenarios.

It is worth emphasizing that the wave-particle energy transfer mechanism examined in this simple Landau damping setup forms the fundamental building block of dissipation in collisionless plasma turbulence. In turbulent plasmas, electromagnetic fluctuations cascade to smaller scales until they can efficiently transfer energy to particles through resonant interactions, precisely the process we are optimizing our closure to represent. While the turbulent case involves a spectrum of waves rather than a single mode, and occurs in a multi-dimensional setting with complex nonlinear interactions, the basic physics of field-to-particle energy transfer remains the same. Therefore, this one-dimensional calibration provides a crucial foundation for our subsequent optimization in more complex turbulent scenarios.

\subsection{Kelvin-Helmholtz Instability}\label{sec:khi}

Having established a preliminary closure parameter from the Landau damping case, we next examine the Kelvin-Helmholtz instability (KHI), which introduces critical multi-dimensional and plasma mixing dynamics absent in the previous test. The KHI represents an intermediate step between the simple linear physics of Landau damping and the full complexity of turbulent cascades, combining large-scale fluid-like behaviour with the emergence of smaller-scale structures where kinetic effects become important.

\begin{figure}
  \centering
  \includegraphics[width=\textwidth]{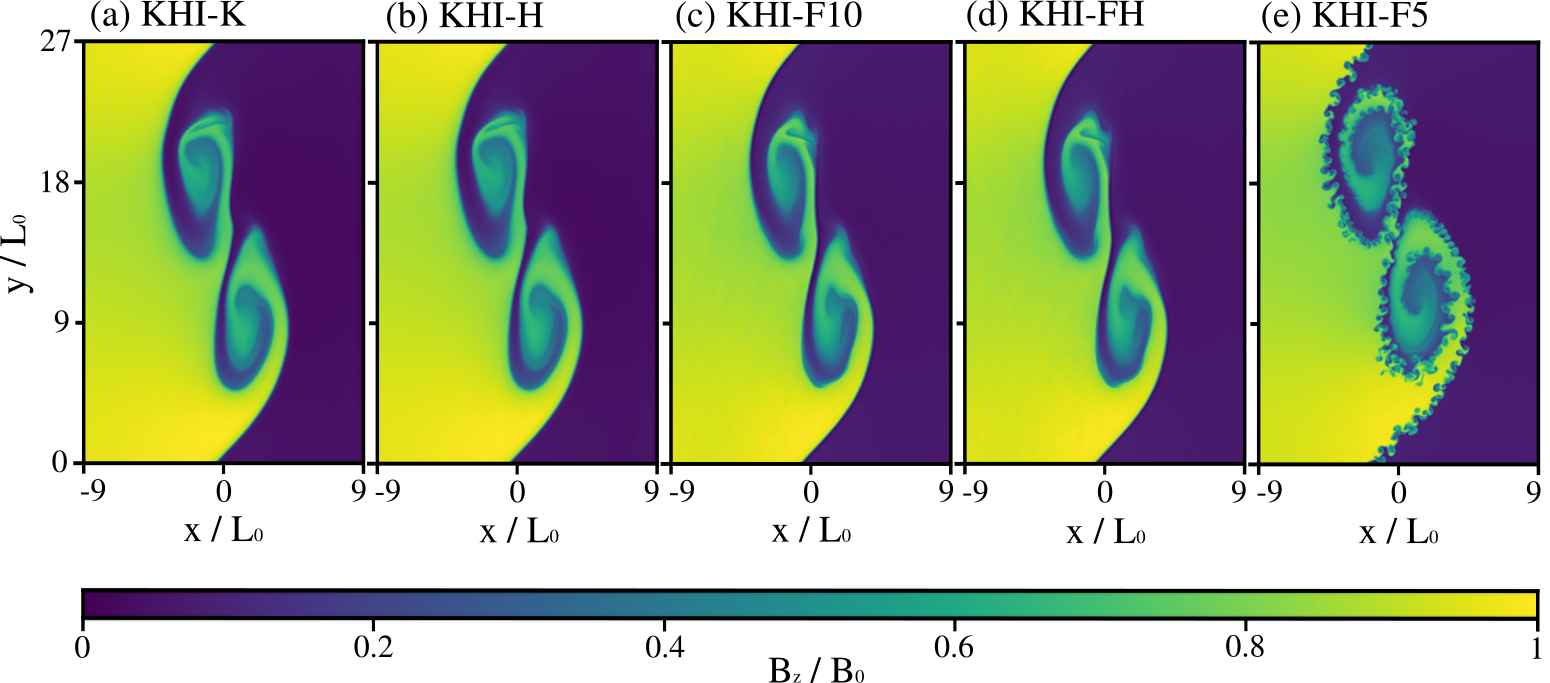}
  \caption{$B_z/ B_{0}$ as a function of space in Kelvin-Helmholtz Instability simulations with different models, at $t = 200 \omega_{p,e}^{-1}$: fully kinetic Vlasov simulation (panel a); hybrid simulation, with Vlasov ions and fluid, 10-moment electrons (panel b); two fluid 10-moment simulation (panel c); two fluid simulation, with 10-moment ions and 5-moment electrons (panel d); two fluid, 5-moment simulation (panel e). The colour indicates the strength of the out-of plane magnetic field $B_z$ normalized to the initial field strength $B_0$. Horizontal and vertical axes show position normalized to the half-width of the initial shear layer $L_0$.}
  \label{fig:khi_Bz}
\end{figure}

This test case serves multiple purposes in our closure study. First, it allows us to evaluate how well our Landau damping-optimized closure translates to multi-dimensional settings with more complex energy dissipation processes, possibly operating at multiple scales. Second, the natural development of vortices and secondary instabilities during KHI evolution tests the closure's ability to handle strongly non-equilibrium scenarios. Third, the KHI setup provides an excellent framework for comparing our hierarchy of models before proceeding to fully developed turbulence. The KHI has already been used as a test case for model comparison in~\citet{henri2013nonlinear}, where MHD, two fluid 5-moment and fully kinetic PIC simulations of KHI are compared. It is observed there that the different models compare well in the linear stage of the instability, when vortices develop at larger scales (tens of ion skin depth), while differences emerge in the non-linear stage, when vortices roll up and smaller-scale dynamics, eventually reaching kinetic scales, appear. 
In this section, we employ all five models described in our methodology, see figure~\ref{fig:khi_Bz}.  

The quantity we depict in figure~\ref{fig:khi_Bz} is the out-of-plane magnetic field $B_z$, normalized to the initial field strength $B_0$ and at time $t = 200\,\omega_{p,e}^{-1}$. We use $k_{0,e}\,d_e = 100$ and $k_{0,i}\,d_i = 10$ for the electron and ion closures in the 10-moment models for the respective species. We notice, however, that the specific values of the parameters $k_{0,e} d_e$ and $k_{0,i} d_i$ do not affect simulation evolution significantly: the crucial ingredient seems to be the possibility to develop agyrotropies and anisotropies (i.e., evolution of the full pressure tensor), rather than the specific value of the closure parameter, see discussion below.  

By comparing the Vlasov and hybrid simulations, panel (a) and (b) in figure~\ref{fig:khi_Bz}, we preliminary observe that the electron model does not affect vortex position and shape: this is consistent with the fact that the KHI is a shearing instability and the fluid velocity in a plasma is essentially the ion velocity. This observation is also supported by the fact that also panel (c) and (d) (10-moment or 5-moment electrons, 10-moment ions) are essentially identical: electron dynamics do not affect KHI evolution. The model we use to simulate ions, instead, affect the internal shape of the vortex, as evident comparing panel (b) and (c), i.e., Vlasov or 10-moment ions: in the presence of 10-moment ions, higher wavenumber (smaller scale oscillations) appear inside the larger scale vortices. When describing ions with a 5-moment model as well, panel (e), we observe the development of even smaller scale vortices due to secondary instabilities at the boundary layers: since heat flux dissipation is absent in the 5-moment model (see discussion in last paragraph of section \ref{sec:muphy}), anisotropic ions are needed to introduce a high-wavenumber cut-off in the modes unstable to KHI. Our results are consistent with previous work.~\citet{cerri2013extended} argument that a sheared velocity field introduces anisotropies in the pressure tensor on very fast temporal scales, and that at least the parallel and perpendicular pressure components have to be evolved independently. They call their model, which evolves 6 equations per plasma species (1 for density, 3 for the current, 1 for the parallel and 1 for the perpendicular pressure component), the \textit{extended} two fluid model.~\citet{del2016pressure} investigated initial condition configurations for the simulation of the KHI. They found that velocity shears introduce not only anisotropies, but also agyrotropies in the VDF, hence the full pressure tensor has to be evolved. Electrons are treated as a massless fluid, supporting our results that electron dynamics are not essential in KHI evolution.  
This model comparison shows that accurate simulations of plasma mixing dynamics in magnetized plasmas requires a 10-moment description, to avoid introducing unphysical wavenumbers in the simulation and to reproduce faithfully the anisotropies and agyrotropies in the VDF introduced by shear flows. This is particularly important in turbulence simulations, addressed in section~\ref{sec:biskamp-welter}, where we want an accurate description of spectra across a wide range of wavenumbers and in the presence of shearing motions. A 10-moment model for the ions, with appropriate closure, appears sufficient to reproduce the main aspects of plasma mixing at a boundary.

\subsection{Decaying Turbulence} \label{sec:biskamp-welter}

Building upon the insights gained from our Landau damping and Kelvin-Helmholtz investigations, we now address the central challenge of this work: identifying the closure parameters that best reproduces key characteristics of decaying turbulence simulations. Decaying turbulence provides an ideal testbed for this work as it naturally generates a broad spectrum of fluctuations across scales, from MHD-like behaviour at large scales to kinetic physics at small scales. This multi-scale energy cascade, and particularly the transition region between fluid and kinetic regimes, is precisely where our Landau-fluid closure must perform accurately. Our choice of $k_{0,e}^o$ and $k_{0,i}^o$ will be guided by the attempt to match at best the spectra from the reference Vlasov simulations, and to reproduce the spatial structure of electron and ion current, as well as the divergence of the electron and ion heat fluxes in the hybrid and 10-moment simulations.

We employ the Biskamp-Welter setup described in section \ref{sec:setup}, which initializes low-correlation velocity and magnetic fluctuations that rapidly develop into complex turbulent structures through nonlinear interactions. Unlike forced turbulence, this decaying configuration allows us to observe how different models handle the natural evolution of the cascade, with particular attention to the transition region from fluid-like behaviour (above the ion skin depth) to the complex dynamics that emerge at kinetic scales.

\begin{figure}
  \centering
  \includegraphics[width=.7\textwidth]{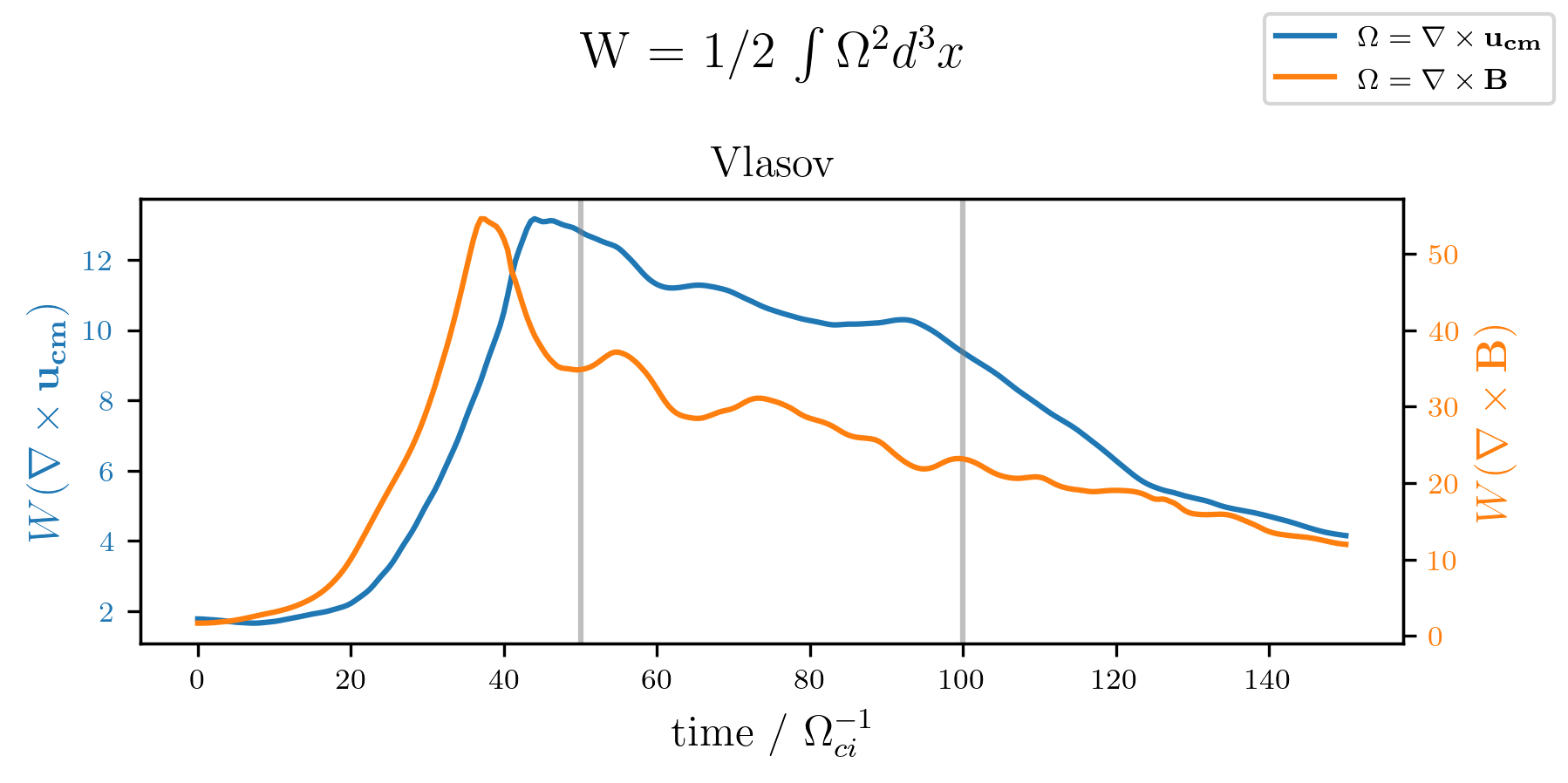}\vspace{-1em}
  \caption{Enstrophy evolution as a function of time in the fully kinetic
    Vlasov simulation for the center of mass velocity and the magnetic field.}
  \label{fig:enstrophy}
\end{figure}

We proceed through distinct phases: First, we evaluate varying electron closure parameters $k_{0,e}$ in hybrid simulations against a fully kinetic reference simulation to establish the optimal $k_{0,e}^o$ for electron dynamics, where the superscript stands for ``\textit{o}ptimal''. Subsequently, we perform 10-moment fluid simulations maintaining the established $k_{0,e}^o$ for electrons while varying the ion parameter $k_{0,i}$ to determine the optimal $k_{0,i}^o$ for ion dynamics. 
To identify the most suitable time range for spectral analysis, we examine the temporal evolution of the enstrophy-type quantities $\int |\nabla \times \vec{u_{cm}}|^2 d^3x$ and $\int |\nabla \times \vec{B}|^2 d^3x$ for the center-of-mass velocity $\vec{u}_{cm}=\frac{m_e \vec{u}_e + {m_i} \vec{u}_i}{m_i+m_e}$ and magnetic field $\vec B$ in the Vlasov simulation, in figure~\ref{fig:enstrophy}.  Starting from the smooth Biskamp–Welter initial condition, strong, isolated current sheets form with nearly exponential growth~\citep{friedel-grauer-marliani:1997} up to time $t \approx 35 \Omega_i^{-1}$. After this initial phase, the current sheets become unstable and turbulence develops.

We select the interval $t \in [50, 100]\,\Omega_i^{-1}$ for our spectral analysis, corresponding to the early decay phase where nonlinear interactions remain significant while avoiding the initial transient dynamics. This interval captures the fully developed turbulent state where the differences between kinetic and fluid models become most apparent, providing an ideal window for comparing spectral characteristics.  The runs we examine are summarized in Table~\ref{tab:BW-runs}. For each run, we report the run ID and the model used. Run K is the fully kinetic run of reference. The runs marked as H are hybrid runs, where ions are simulated with the Vlasov model and electron with the 10-moment model and the closure described in equation~\ref{eq:tenmoment_closure}. The $k_{0,e}$ we use in each run is listed in the corresponding row. The ``optimal'' $k_{0,e}^o$, chosen as described in the rest of the section, is marked in bold.  The F10 runs employ the 10-moment model for both electrons and ions. We close the electrons using the optimal $k_{0,e}^o$ selected in the hybrid runs. The different $k_{0,i}$ we use are listed in the corresponding lines. The ``optimal'' $k_{0,i}^o$ is again marked in bold.

\begin{table}
  \centering
\begin{tabular}{l@{\hskip 10pt}|@{\hskip 10pt}l@{\hskip 10pt}|@{\hskip 10pt}c@{\hskip 10pt}|@{\hskip 10pt}c}
\textbf{Run ID} & \textbf{Model} & $\mathbf{k_{0,i} d_i}$ & $\mathbf{k_{0,e} d_e}$  \\\hline
BW-K & Kinetic & -- & -- \\[5pt]
BW-H-1 & Hybrid & -- & 2.0 \\
BW-H-2 & Hybrid & -- & 20 \\
BW-H-3 & Hybrid & -- & \textbf{200} \\
BW-H-4 & Hybrid & -- & 2\,000 \\[5pt]
BW-F10-1 & Fluid10 & 2.0 & 200 \\
BW-F10-2 & Fluid10 & \textbf{20} &  200 \\
BW-F10-3 & Fluid10 & 200 & 200 \\
BW-F10-4 & Fluid10 & 2\,000 & 200 \\
\end{tabular}
\caption{List of the turbulence runs examined in this work. For each run, we provide a run ID (BW stands for Biskamp-Welter), the model type (kinetic, hybrid or two-fluid 10-moment) and, when needed, the $k_0$ value used for the heat flux closure, normalized to the respective skin depth. The $k_0$ value that compares best with the kinetic benchmark is marked in bold.}
\label{tab:BW-runs}
\end{table}

\begin{figure}
  \centering
  \includegraphics[width=\textwidth]{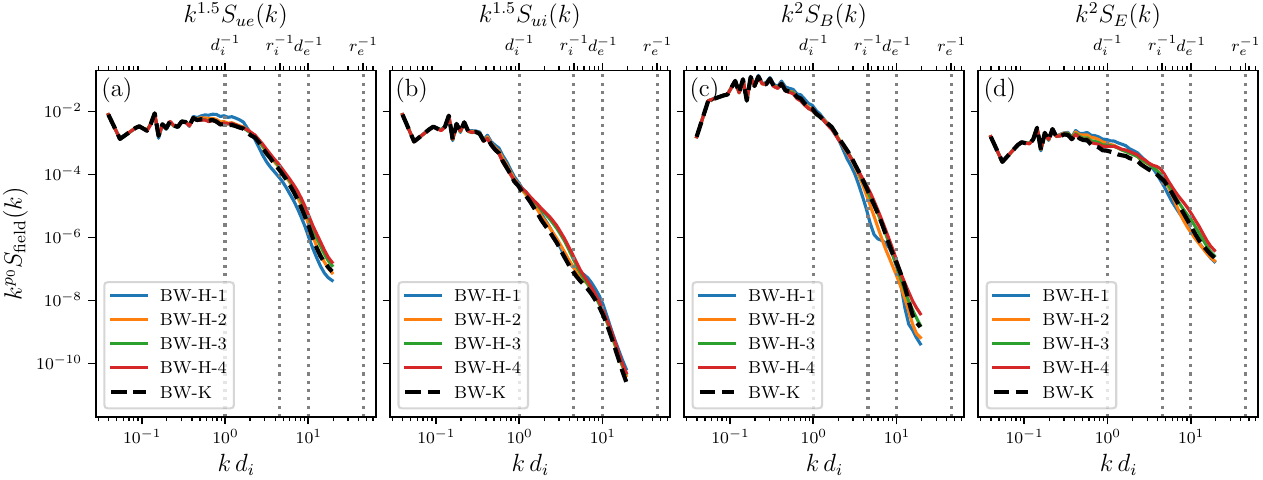}
  \caption{Compensated spectra for the hybrid simulations with varying $k_{0,e}$, compared with the Vlasov reference: (a) electron velocity, (b) ion velocity, (c) magnetic field, (d) electric field spectra. The different hybrid simulations are marked in colours. The Vlasov reference is depicted in dashed line.}
  \label{fig:BW-spectra-hybrid}
\end{figure}

We start examining the hybrid simulations. In figure  \ref{fig:BW-spectra-hybrid}, we depict the electron velocity, ion velocity, magnetic field and electric field spectra, as a function of the wavenumber $k\;d_i$, for the Vlasov, BW-K, simulation and for the hybrid simulations, run with the different $k_{0,e} d_e$. The spectra are compensated by multiplying them by $k^{po}$ where $k^{po}=1.5$ for the velocity spectra and $k^{po}=2$ for the field spectra. In all cases, we observe a remarkable similarity between the hybrid spectra, notwithstanding the different $k_{0,e}$ closure, and the Vlasov reference.
  
\begin{figure}
  \centering
  \includegraphics[width=\textwidth]{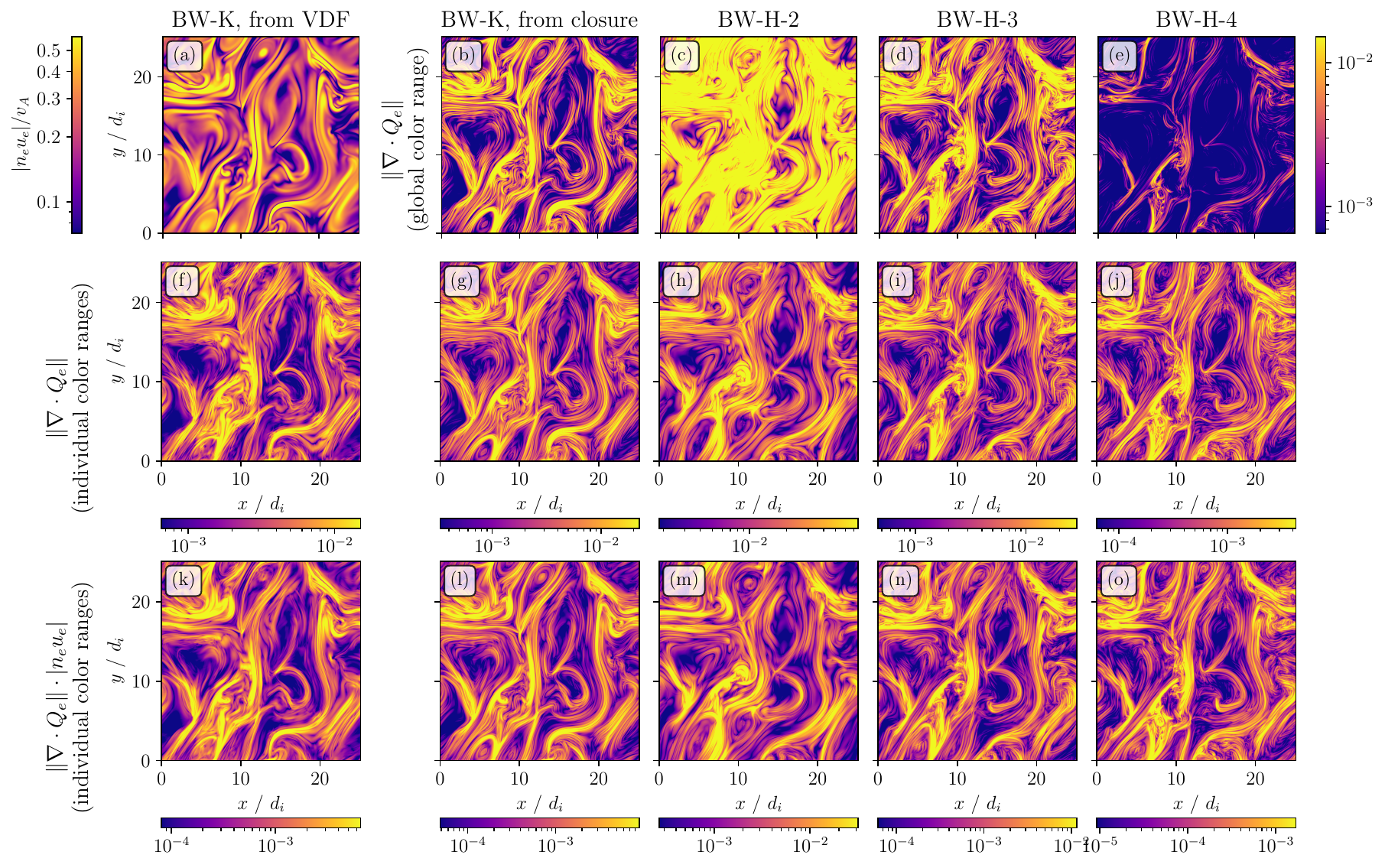}\\
  \caption{Comparison of 2D colormaps at time $t=100\,\Omega_{c,i}^{-1}$ showing:
  Panel (a): $|j_e|$ from the Vlasov reference simulations. Panel (f): $\|\nabla\cdot Q_e \|$ calculated from the distribution function of the Vlasov simulation; Panel (g): calculated with equation~\eqref{eq:tenmoment_closure} using moments from the Vlasov simulations. By matching amplitude with (f), we a posteriori determine $k_{{0,e}}\,d_e= 201$; $\|\nabla\cdot Q_e \|$ from hybrid simulations using (c, h) $k_0 d_e= 20$, (d, i) $k_0 d_e= 200$ and (e, j) $k_0 d_e= 2000$. In (g)-(j), we adjust the color scale in every panel. (b)-(e) show the same data with color scales set to the range of the kinetic ``ground truth'' (f). (k)-(o) depict $\| \nabla\cdot Q _e\| \cdot |j|$, with $\|\nabla\cdot Q_e \|$ calculated as in the row above.  }
  \label{fig:divq_hybrid}
\end{figure}

In figure~\ref{fig:divq_hybrid} we plot several fields in the $xy$ plane at time  $t=100\,\Omega_{c,i}^{-1}$. The spectral norm of the closure expression for each point in space is calculated as~$\|\nabla \cdot Q_s\|(x,y) = \sigma_{\textup{max}}((\nabla \cdot Q_s)(x,y))$, where $\sigma_{\textup{max}}()$ returns the largest singular value of the argument.
First, in the first row, panel (a), we depict the magnitude of the electron current density $\mathbf{j}_e= n_e \mathbf{u}_e$ in the reference Vlasov simulation, to highlight the evolution of the turbulent patterns in the simulations. In panel (f), $\|\nabla \cdot Q_e\|$ is calculated directly from the velocity distribution function in the Vlasov simulations: this is the ``ground truth'' we strive to reproduce by selecting $k_{0,e}$ appropriately. In panel (g), as a comparison, we depict the absolute value of the heat flux calculated \textit{from the moments of the Vlasov simulation} using equation~\eqref{eq:tenmoment_closure} and a $k_{0,e}$ value, $k_{0,e} d_e= 201$, obtained by comparison with the results in panel (b). Comparing panel (f) and panel (g), we are reassured that equation~\eqref{eq:tenmoment_closure} is a good choice for our closure. We expect similar, but not identical, results when plotting the closure obtained from equation~\eqref{eq:tenmoment_closure} for the hybrid simulations (panels (h), (i), (j)), because the lower moments used to calculate the closure are affected by the closure itself. In panels (h), (i), (j) we depict the closure obtained from equation~\eqref{eq:tenmoment_closure}, with the $k_0 d_e$ value indicated in the caption. In the center row of figure~\ref{fig:divq_hybrid}, we rescale the individual plots to \textit{ad hoc} colour scales: the aim is to highlight how the turbulent patterns change with the different $k_{0,e}$. In the top row, panels (b), (c), (d), (e) we use for all plots the same colour range as the kinetic benchmark (f): we want to identify which $k_{0,e}$ value delivers magnitude of the heat flux closer to the kinetic benchmark. As one could expect from equation~\eqref{eq:tenmoment_closure}, higher $k_0 d_e$ values result in lower $\|\nabla \cdot Q_e\|$ magnitudes.  After analysing figure~\ref{fig:BW-spectra-hybrid} and figure~\ref{fig:divq_hybrid}, we choose $k_{0,e} d_e= 200$ as our best fit. We however remark that both spectra and turbulent patterns do not change significantly with varying $k_{0,e}$ in the hybrid simulations. In figure~\ref{fig:divq_hybrid} panels (k) to (o) we depict the electron $\|\nabla \cdot Q_e\| \cdot |j_e|$, with $\|\nabla\cdot Q_e\| $ in the different panels calculated as in panels (f) to (j). By comparing the different panels with corresponding plots for the electron current (see panel (a) for the electron current density in the Vlasov simulation), we see that the peaks of $\|\nabla\cdot Q_e\| $ are well aligned with the areas of maximum current density. This suggests that it is particularly critical to correctly capture heat flux closures in correspondence of the current density peaks, if one intends to reproduce fundamental turbulence properties such as intermittency~\citep{wan2012intermittent}.

\begin{figure}
  \centering
  \includegraphics[width=\textwidth]{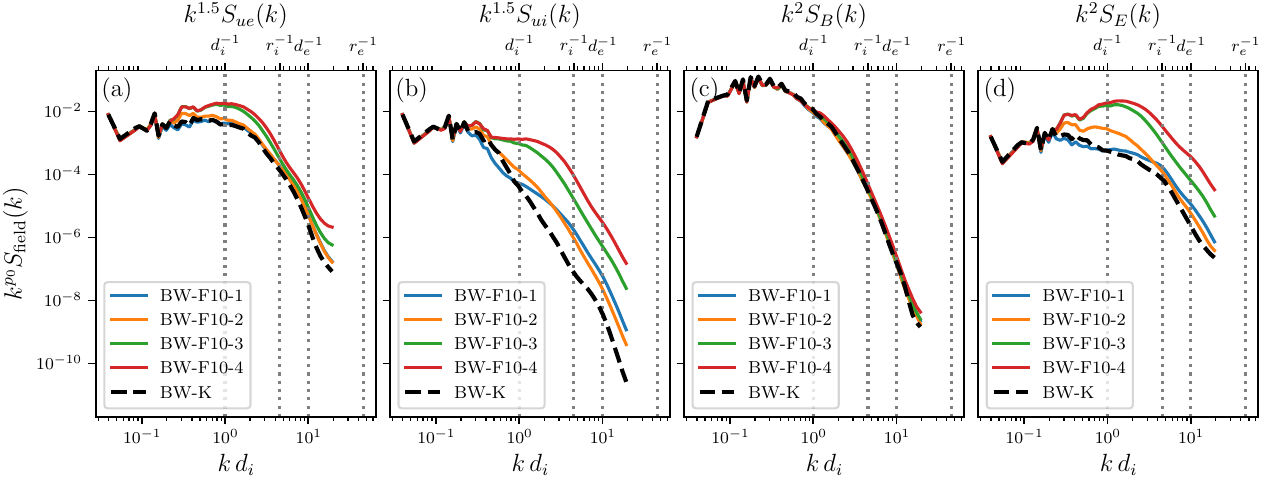}
  \caption{Compensated spectra for the fluid simulations with $k_{0,e} d_e = 200$ and with varying $k_{0,i}$, compared with the Vlasov reference: (a) electron velocity, (b) ion velocity, (c) magnetic field, (d) electric field spectra. The different two fluid simulations are marked in colors. The Vlasov reference is depicted in dashed line.}
  \label{fig:BW-spectra-fluid}
\end{figure}

After having identified $k_{0,e}^o d_e= 200$ as the the optimal closure parameter for our hybrid simulations, we repeat the same exercise with the two fluid 10-moment simulation series in Table~\ref{tab:BW-runs}. As  for the hybrid simulations, we plot in figure~\ref{fig:BW-spectra-fluid} the compensated electron velocity, ion velocity, magnetic field and electric field spectra as a function of the wavenumber. We depict results for the Vlasov, K, simulation and for the 10-moment simulations, run with  $k_{0,e}= k_{0,e}^o$ for the electrons and the different  $k_{0,i}$. for the ions. We observe that, in this case, the choice of the closure parameter affects quite significantly the different spectra, with the exception of the magnetic field spectra.

\begin{figure}
  \centering
  \includegraphics[width=\textwidth]{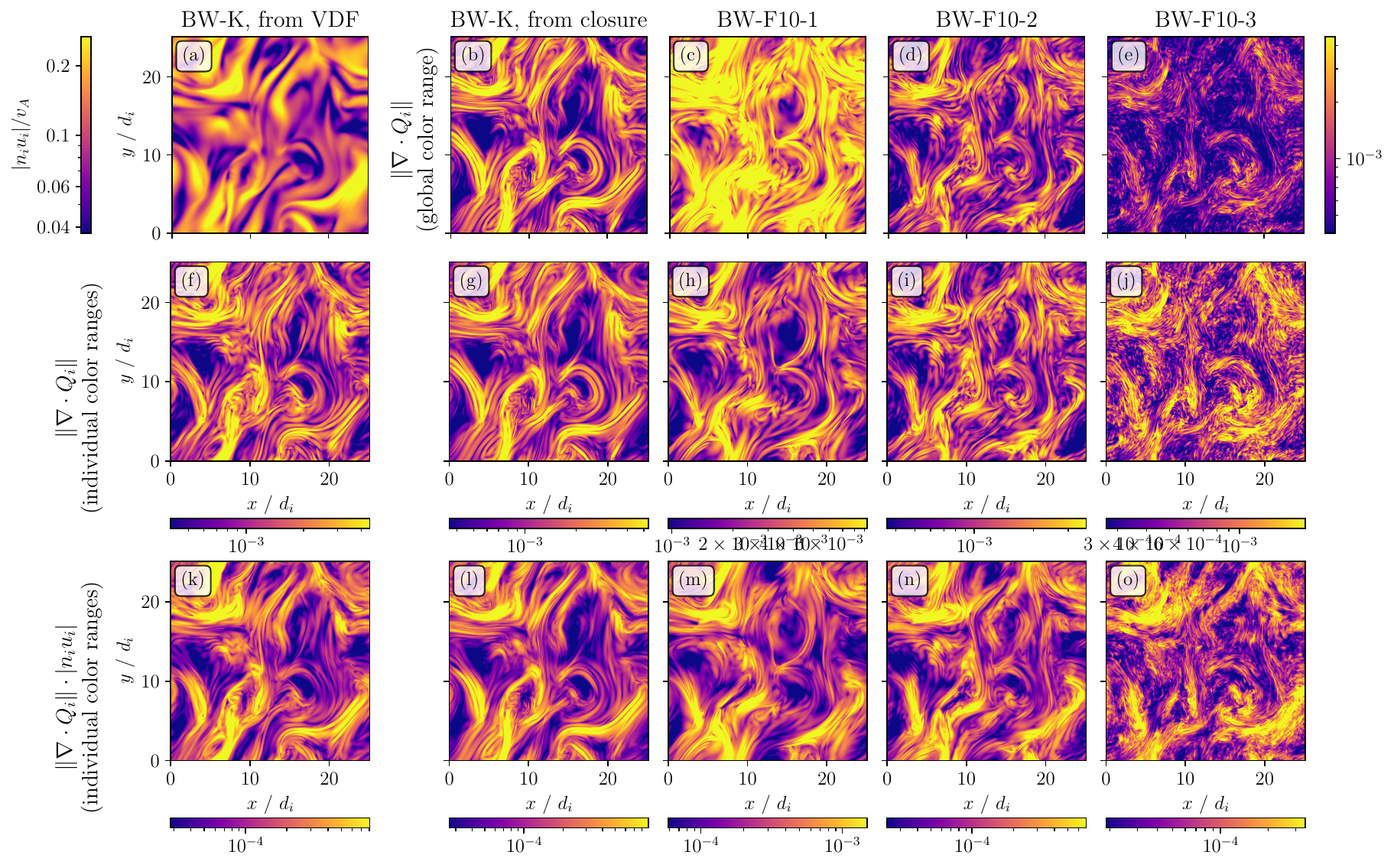}\\
  \caption{Comparison of \textit{ion} heat flux divergence $\| \nabla\cdot Q _i\|$ at time $t=100\,\Omega_{c,i}^{-1}$ for ten-moment simulations with $k_{0,e} d_e = 200$ and varying $k_{0,i}$: the ion closures use (c,h,m) $k_0 d_i= 2$, (d,i,n) $k_0 d_i= 20$ and (e,j,o) $k_0 d_i= 200$. Panels compositions is the same as in figure \ref{fig:divq_hybrid}.}
  \label{fig:divq_10mom_i}
\end{figure}

\begin{figure}
  \centering
  \includegraphics[width=\textwidth]{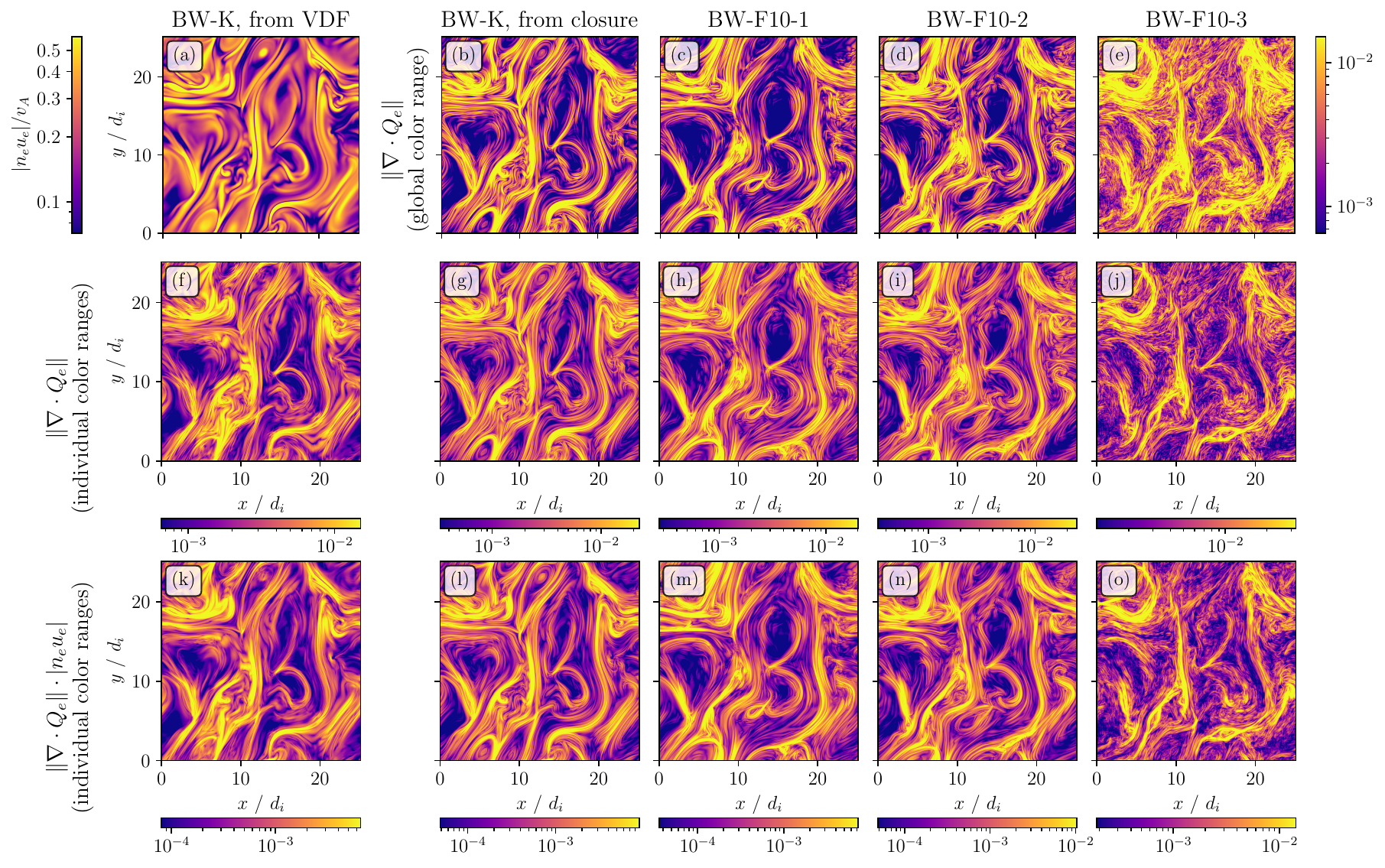}\\
   \caption{Comparison of \textit{electron} heat flux divergence $\| \nabla\cdot Q _e\|$ at time $t=100\,\Omega_{c,i}^{-1}$ for ten-moment simulations with $k_{0,e} d_e = 200$ and varying $k_{0,i}$: the ion closures use (c,h,m) $k_0 d_i= 2$, (d,i,n) $k_0 d_i= 20$ and (e,j,o) $k_0 d_i= 200$. Panels compositions is the same as in figure \ref{fig:divq_hybrid}.}
  \label{fig:divq_10mom_e}
\end{figure}

Figures~\ref{fig:divq_10mom_i} and \ref{fig:divq_10mom_e} highlight the effects of closures applied to the ion species to both ion (figure \ref{fig:divq_10mom_i}) and electron (figure \ref{fig:divq_10mom_e}) quantities. 
Note that we only vary the ion closure, the variation in figure \ref{fig:divq_10mom_e} shows the effect of the ion closure on the electron species. First, in panel (a) we depict the electron and ion current density respectively. We observe that, as expected, electron current structures develop at smaller scales than ions. Panel (f) and (g) depict $\|\nabla\cdot Q_s\| $ from Vlasov simulations, with $s$ the particle species. In panel (f) we depict the ``true'' $\|\nabla\cdot Q_s\| $, calculated from the distribution function, in panel (g) equation~\eqref{eq:tenmoment_closure}, calculated using the moments from the Vlasov simulations. Quite surprisingly, we observe in figure~\ref{fig:divq_10mom_i} that the ion $\|\nabla\cdot Q_i\| $ exhibit signatures at smaller scales than those of the ion current density. In panels (c) to (e) and (h) to (j), we depict the $\|\nabla\cdot Q_s\| $ closure, calculated following equation~\ref{eq:tenmoment_closure} from the two fluid 10-moment simulations reported in the respective column titles. The $k_{0,i} d_i$ for the ions is $k_{0,i} d_i=2, 20, 200$, while for the electron we keep it fixed at $k_{0,e}^o d_e= 200$. As before in figure~\ref{fig:divq_hybrid}, we use in the first row the same colour scale as the ``ground truth", individual colour scales in the second row. 
Comparing the fluid ion closures with the Vlasov ``ground truth'' we observe, this time rather starkly, the effects that the choice of the $k_{0,i}$ parameters has on the formation of patterns at different scales: The higher the $k_{0,i} d_i$ (i.e., the lower the $\nabla\cdot Q_i$), the more small-scale signatures emerge in the fluid closures. Comparing the electron closures from the fluid simulations with the Vlasov ``ground truth'' and the respective ion closure, we notice the effect of large-to-small scale coupling: the high magnitude signatures in the electron closures resemble more closely the respective ion signatures rather than the Vlasov electron ``ground truth''. This is particularly evident in the last column, $k_{0,i} d_i= 200$. The rather high value of $k_{0,i} d_i$ results into unphysical small scale structures in the ion closure: the same small-scale structures are visible in the electron closure in figure~\ref{fig:divq_10mom_e} as well. Furthermore, both the $k_{0,i} d_i= 2$ and $k_{0,i} d_i= 20$ electron closures (figure~\ref{fig:divq_10mom_e}, panels (c) and (d) ) exhibit a specific ``double-arcade'' pattern at $10 \lesssim x/ d_i \lesssim 20$; $y/ d_i \lesssim 2.5$. The same pattern appears in the respective ion closures, figure~\ref{fig:divq_10mom_i}, but not in the ``ground truth'' Vlasov electron heat flux, figure~\ref{fig:divq_10mom_e}, panel (f), nor in the depiction of equation~\ref{eq:tenmoment_closure} calculated with moments from the Vlasov simulations in figure~\ref{fig:divq_10mom_e} panel (g): it must have been driven by  the selected ion closure. 

In figures \ref{fig:divq_10mom_i} and \ref{fig:divq_10mom_e} panels (k) to (o) we depict $\|\nabla\cdot Q_s\| \cdot |j_s|$, with $\|\nabla\cdot Q_s\| $ calculated as in figure \ref{fig:divq_hybrid} and $s=i,\:e$ respectively. Also here, we observe that the peaks of $\|\nabla\cdot Q_s\| $ are well aligned with those of $  |j_s|$. 

Analysing together figures \ref{fig:BW-spectra-fluid}, \ref{fig:divq_10mom_i}, and \ref{fig:divq_10mom_e} we choose $k_{0,i} d_i= 20$ as the optimal closure parameters for the ions. 

\begin{figure}
  \centering
  \includegraphics[width=\textwidth]{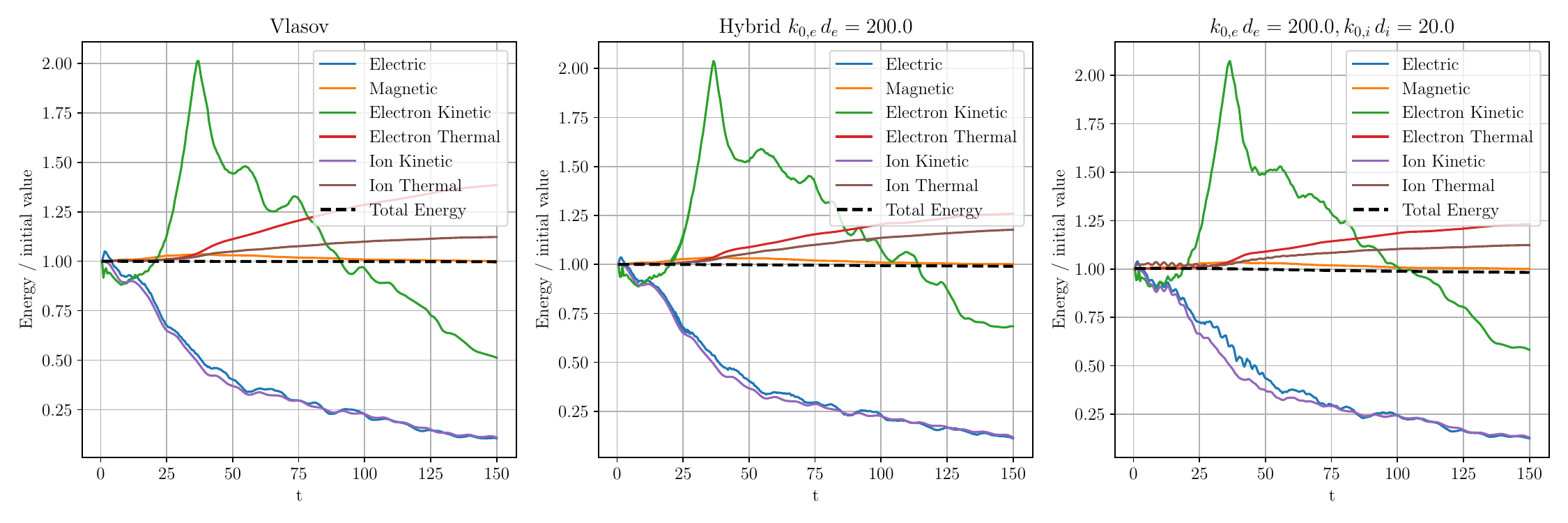}\\
  \includegraphics[width=\textwidth]{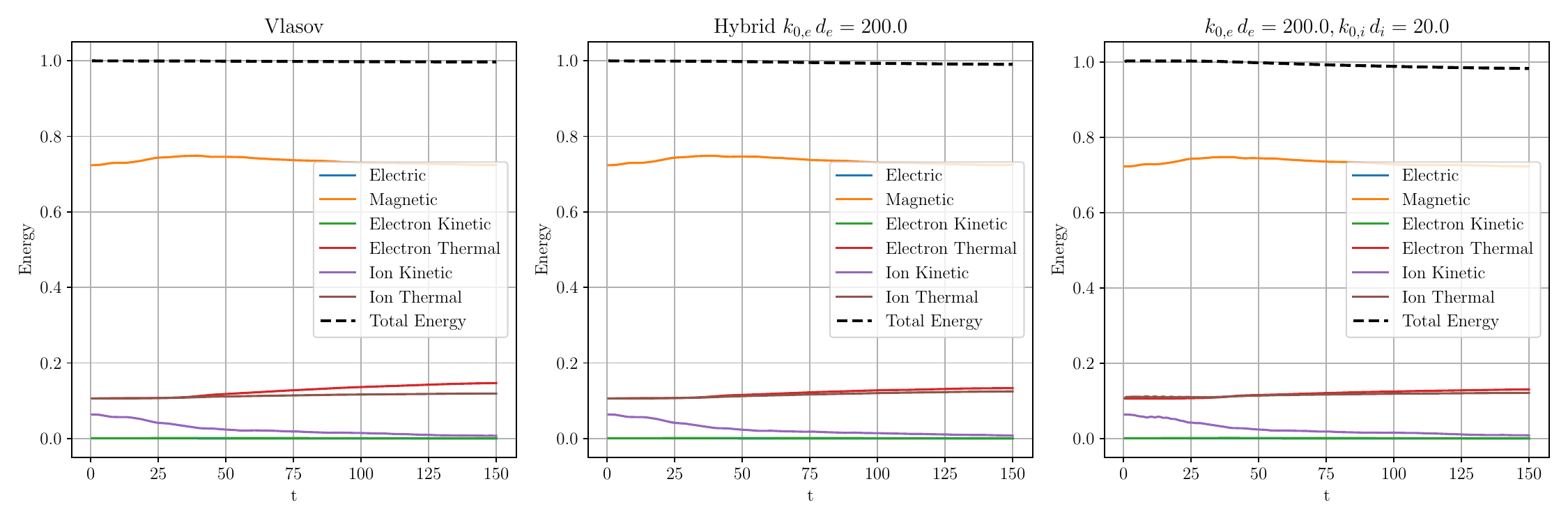}\vspace{-1em}
  \caption{Energy evolution for the Vlasov simulation (first column), ``best'' hybrid simulation, with $k_{0,e}^0 d_e = 200$ (second column) and ``best'' two fluid 10-moment simulation, with with $k_{0,e}^0 d_e = 200$ and $k_{0,i}^0 d_i = 20$ (third column).
    Top row: Total energy and each component is normalized to their respective initial value.
    Bottom row: Total energy and each component is normalized to initial value of total energy.}
  \label{fig:energy}
\end{figure}

In figure~\ref{fig:energy} we depict the energy evolution as a function of time for three simulations: the Vlasov simulation (first column), the ``best'' hybrid simulation, with $k_{0,e}^0 d_e = 200$ (second column) and the  ``best'' two fluid 10-moment simulation, with  $k_{0,e}^0 d_e = 200$ and $k_{0,i}^0 d_i = 20$ (third column). In the first row, the total energy and the different energy components (electric, magnetic, electron kinetic and thermal, ion kinetic and thermal) are normalized to their initial respective value; in the second row, to the total initial energy. Comparing the three simulations, we observe excellent energy conservation in all cases (minimal energy decrease is observed in the two fluid simulation), and also good agreement between the evolution of the different energy components between the Vlasov, hybrid and two fluid cases. In the first row of figure~\ref{fig:energy} we observe some difference between the electron kinetic and thermal energy evolution in the Vlasov simulation, on the one hand, and hybrid and two fluid simulations on the other: At the end of the Vlasov simulation, the electrons exhibit lower kinetic energy than in the hybrid and two fluid simulations, but are hotter. We observe however in the second row that both the electron thermal and, in particular, the electron kinetic energy components are rather low, in percentage, with respect to the total energy content of the simulation. This is the reason why this different distribution of the electron energy seems to impact minimally on the Vlasov and hybrid spectra, which are remarkably similar (see figure~\ref{fig:BW-spectra-hybrid}). On the other hand, comparing the Vlasov, hybrid and two fluid energy evolution, we can conclude that the difference in spectra between the two fluid and Vlasov case in figure~\ref{fig:BW-spectra-fluid} is not related to different energy distribution among the different energy components, but rather to different energy distribution between different \textit{scales}: In figure~\ref{fig:BW-spectra-fluid} we see that higher $k_{0,i} d_i$'s result in higher energy at high wavenumber for the electron and ion velocity spectra, and for the electric field spectra. In the energy plot, however, no significant difference between the Vlasov and two fluid simulations is visible for the electron field and for electron and ion kinetic energy.

\section{Discussion and Conclusions}\label{sec:discussion}

This work is motivated by one of the most challenging aspects of computational plasma physics, and computational space physics in particular: Since plasmas are inherently multi-scale systems, simulations need to accommodate, together, the large box sizes and temporal durations needed to capture large-scale, slow processes and the small-scale, fast, kinetic processes that critically influence energy dynamics and hence even large scale evolution. We investigate here two-fluid 10-moment models as a compromise between fully kinetic and MHD models, which have the respective drawbacks of being either extremely computationally expensive or inaccurate in the modelling of kinetic processes.
10-moment methods evolve velocity moments up to the pressure tensor included, and therefore need a closure for the divergence of the heat flux tensor. The choice of the closure is essential in reproducing the effects at large scales of kinetic scale energy transport processes, at least for some of the test cases we examine.

In this paper we have focused on decaying turbulence, a problem where, on the one hand, one needs large separation in wavenumber between the scale of energy injection and that of energy dissipation and, on the other hand, energy dissipation at kinetic scales has to be properly modelled for its effects in determining energy spectra at intermediate and kinetic scales. We have chosen a specific local Landau fluid closure, equation~\eqref{eq:tenmoment_closure}, and we have investigated if specific choices of the closure parameters $k_{0,s}$ for both electrons and ions can deliver spectra and turbulent structures which compare well against reference fully kinetic simulations. Before addressing the decaying turbulence problem, we have tuned our methodology on ``simpler'' problems, i.e. Landau damping and Kelvin--Helmholtz instability, where we could address separately the role of the heat flux closure in wave damping and velocity shear instability, respectively.

The results of our work are the following: First, our investigations on the Landau damping problem shows that 10-moment models with our specific Landau fluid closure and an appropriately chosen closure parameter $k_{0,s}$  reproduce wave damping into the plasmas (damping rate, period of the electric field oscillation) with sufficient accuracy for our intended usage in large scale simulations. This is not a surprise, since Landau fluid closures have been designed to reproduce exactly this process. Second: our Kelvin-Helmholtz simulations show that a ``minimum'' model exists for the simulation of velocity shear-driven instabilities such as the Kelvin--Helmholtz instability. This ``minimum'' model is a 10-moment model for the ions, 5-moment model for the electrons, that allows for the formation of ion anisotropies and agyrotropies triggered by the presence of a velocity shear. Within this model, a wide range of $k_{0,i}$ closure parameters for the ions deliver comparable results. In contrast, when a 5-moment model is used for the ions as well, high-wavenumber spurious oscillations are observed in the simulations. This result is of extreme interest for a wide range of physical problems, from turbulence simulations to global magnetospheric simulations, where velocity shear instabilities occur. Third, our decaying turbulence simulations highlight the role of the ions in energy transport from fluid-like to kinetic scales: while in 10-moment hybrid simulations (fluid electrons, kinetic ions) energy spectra for both fields and particles are well reproduced notwithstanding the choice of the electron closure parameters, the choice of the closure parameter for the \textit{ions} is more critical. When testing multiple $k_{0,i}$  closure parameters for the ions, we aim at matching against the reference Vlasov simulations both the energy spectra and the magnitude and structures in the divergence of the heat flux calculated in the 10-moment simulations from lower moments with equation~\eqref{eq:tenmoment_closure}. The $k_{0,i}$  value we consider ``optimal'' after comparison with the Vlasov reference, $k_{0,i}\,d_i=  20$, satisfies both these requirement. Lower and higher $k_{0,i}$ values tend to respectively over- and under-estimate heat flux divergence, as one could expect from the closure we use. Interestingly, higher $k_{0,i}$ values also results in the appearance of high-wavenumber spurious oscillations in the calculated heat flux divergence for both the ions and the electrons, showing that the ions are essential in transmitting energies from the fluid-like to the kinetic scales. This conclusion is further strengthened by the observation that structures in the electron closure  from 10-moment simulations appear to map  directly to similar structures in the ion closures, which are absent in the Vlasov ``ground-truth''.

Further considerations can be derived from our work. First, we have demonstrated the robustness of Landau-fluid closures with respect to some of the assumptions under which they are derived. Landau fluid closures are derived under the assumption that the plasma is close to Local Thermodynamic Equilibrium (LTE). However, they still perform quite well in our Kelvin--Helmholtz and decaying turbulence test cases (which, as most of the plasma processes of interest, are far from LTE) as long as appropriate closure parameters $k_{0,s}$ are found. Second, the closure parameters $k_{0,s}$ that we have identified as ``optimal'' in all our test cases not only reproduce well the large-scale consequences of energy dissipation, but also are in the range of the spatial scales where we expect dissipation processes to become of relevance for the different plasma species~\citep{ng2020improved}. This confirms that we can interpret the inverse of the $k_{0,s}$ parameter as a characteristic length for heat transport and gives a hint for how the parameter should be chosen. Third, we have highlighted the importance of appropriately approximating heat flux closures especially in regions of high current densities, which are associated with intermittent and dissipative behaviour. This hints to the possibility of using clustering methods in simulations (see e.g.~\citet{kohne2023unsupervised, Donaghy_Germaschewski_2023}) to identify dissipative regions, and then design heat flow closures, whereby the training process is conditioned on regions with high dissipation.

We acknowledge a number of limitations in our current investigations, which will constitute the basis for future work. First, we have demonstrated that optimal closure parameters exist under very specific plasma conditions, without verifying how well these parameters perform when plasma or simulation parameters change. Examples of the former are plasma betas, parallel to perpendicular temperature ratio; of the latter, ion to electron mass ratio, box size. Proving that the same closure parameters can be used under different plasma conditions or that effective closure parameters can be derived as a function of plasma conditions is the focus of ongoing investigation. Second, and perhaps most important: another assumption behind the Landau-fluid closure we used is that transport is isotropic. This is the case in the simulations we have presented here, but very much not in the space physics systems we ultimately target, such as the solar wind, where parallel and perpendicular transport are quite different. In future work, we intend to verify if this anisotropy can be incorporated into 10-moment models while retaining our current closure, equation~\eqref{eq:tenmoment_closure}, perhaps using different closure parameters in the parallel and perpendicular direction, or if different closure equations have to be used in the different directions.

\begin{acknowledgments}
  The authors acknowledge support from the German Research Foundation DFG within the Collaborative Research Center SFB1491 and projects 497938371.  We also gratefully acknowledge the Gauss Centre for Supercomputing e.V. (www.gauss-centre.eu) for funding this project by providing computing time through the John von Neumann Institute for Computing (NIC) on the GCS Supercomputer JUWELS at Jülich Supercomputing Centre (JSC) and SuperMUC-NG at the Leibniz Supercomputing Centre.  Computations were conducted on JUWELS/JUWELS-booster, SuperMUC-NG and on the Galileo cluster of the SFB1491.
\end{acknowledgments}

\bibliographystyle{jpp}
\bibliography{bibliography}

\end{document}